\documentclass[12pt]{article}
\usepackage{bm}
\usepackage{amstext}
\usepackage{graphicx}
\begin{document}
\title{Kagome Approximation for $^{3}He$ on Husimi Lattice with Two- and Three-Site Exchange Interactions}
\author{N.S. Ananikian$^1$, V.V. Hovhannisyan$^1$, H.A. Lazaryan$^2$,\\
\begin{footnotesize}\textit{$^1$
A.I. Alikhanyan National Science Laboratory, Alikhanian Br. 2, 0036 Yerevan, Armenia,}\end{footnotesize}\\
\begin{footnotesize}\textit{$^2$
Department of Theoretical Physics, Yerevan State University,
}\end{footnotesize}\\
\begin{footnotesize}\textit{
A. Manoogian 1, 0025 Yerevan, Armenia.}\end{footnotesize} }

\maketitle

%+Abstract
\begin{abstract}
The Ising approximation of the Heisenberg model in a strong magnetic
field, with two-, and three-spin exchange interactions are studied
on a Husimi lattice. This model  can be considered as an
approximation of the third layer of  $^3He$ absorbed on the surface
of graphite (kagome lattice). Using dynamic approach we have found
exact  recursion relation for the partition function. For different
values of exchange parameters and temperature the diagrams of
magnetization are plotted and showed that magnetization properties
of  the model vary from ferromagnetic to antiferromagnetic depending
from the value of model parameters. For antiferromagnetic case
magnetization plateau  at $1/3$ of saturation field is obtained.
Lyapunov exponent for recursion relation are  considered and showed
absents of bifurcation points in thermodynamic limit. The Yang-Lee
zeros are analyzed in terms of neutral fixed points and showed that
Yang-Lee zeros of the model are located on the arcs of the circle
with the radius $R=1$.
\end{abstract}
%-Abstract

%+Contents
%\tableofcontents
%-Contents

\section{Introduction}

Since the spin-1/2 $^3He$
 atoms have     week attractive  potential and light mass the zero-point fluctuations are large. Therefore
helium can solidify only at very high pressures. Unlike usual
solids,   where  the dominant is dipolar magnetic nuclear
interaction in  solid  $^3He$    nucleon-nucleon interaction is
dominant, therefore nuclear spins become ordered at 1 mK (for
usual solids 1~microKelvin)\cite{Roger}. For such systems theory
of magnetism is based on multiple-spin exchange mechanism.

It is important to examine solid and fluid $^3He$ films absorbed on
the surface of
graphite,\cite{graffite,graffite1,graffite2,graffite3} since it is a
typical example of a two-dimensional frustrated quantum-spin
system\cite{frustrated,frustrated1}. The first and second layers of
that system form triangular lattice, while the third one forms a
system of quantum 1/2 spins on kagome
lattice\cite{kagome,kagome1,kagome2}. Both
experimental\cite{experemental,experemental1,experemental2} and
theoretical\cite{theoretical,theoretical1} studies indicate that
three site exchange interaction is dominant in these systems. When
the density of $^3He$ nuclei decreases the magnetic properties of
system changes from ferromagnetic to antiferromagnetic one. This
behavior can be explained in terms of multiple-spin exchange (MSE).
For  high densities the system is fully packed and three-site
exchange interaction is dominant, therefore the system is
ferromagnetic. When density decreases two-site exchange interaction
becomes dominant and magnetic properties of the system change into
antiferromagnetic one.

The magnetic properties were studied in Bose-Einstein condensates of
 ultracold atoms. The dynamical creation of fractionalized halt-quantum vortices
 in Bose-Einstein condensate of sodium atoms have been demonstrated\cite{Josephson}.
  Moreover, there were made the description
 of the Josephson effect in Abelian and non-Abelian  Bose-Einstien
 condensate of alkali elements\cite{Josephson1,Josephson2}.
  Last of them may be realised in superfluid $^3He$ Josephson weak
   link\cite{Josephson3}. The Kosterlitz-Thouless transition with random
    magnetic field\cite{Josephson4} was used to describe marginal metal to
    insulator transition in disordered
graphene\cite{Josephson5}.

 At low temperatures antiferromagnetic quantum atoms of $^3He$ can
exhibit plateaus on the magnetization curve. Beginning at some value
of the external magnetic field (less than saturation field) the
magnetization of the system does not change when the external
magnetic field increase and  system  adsorbs energy without any
change of magnetization. When the external magnetic field reaches a
certain value the magnetization changes its value again.

 This phenomena was theoretically predicted by Hida\cite{hida}
for  a ferromagnetic-ferromagnetic-antiferromagnetic Heisenberg
chain 3CuCl$_2\cdot$2  dioxane compound, which consists of the
antiferromagnetic coupled trimers. For the different models the
magnetization plateaus are predicted to occur in chains, ladders,
within the dynamical and transfer matrix approaches (see
Ref.~\cite{isingpotshezenberg}-\cite{platoes6}).  For two
dimensional systems  appearance of magnetization plateaus   has been
found experimentally. On the triangular lattice a magnetization
plateau was observed at $m/m_s=1/3$ for compounds like
C$_6$Eu\cite{CEu,CEu1},  CsCuCl$_3$\cite{CsCuCl}(see also Ref.
\cite{NH4CuCl3}-\cite{NH4CuCl36}). For kagome lattice also have been
observed magnetization plateau at 1/3 of saturation
field\cite{kagomeplato,kagomeplato1}. Noticeable, that plateaus
appear only for the values of magnetization  which are quantized to
fraction values of the saturation magnetization. This phenomena
 was  theoretically explained by Oshikawa, Yamanaka and  Affleck\cite{Oshikwa}
in 1997.

As mentioned above the third layer of the  $^3He$ films absorbed on
the surface of graphite is kagome lattice. Usually, the
antiferromagnetic kagome lattice is investigated using numerical
simulation\cite{numericalkagome,numericalkagome1,numericalkagome2}.
In this paper  dynamic system approach based on exact recursive
relation for the partition function have been used. The key point of
the approach is the so-called recursive lattice
approximation\cite{bete}, which is a very powerful tool in
investigating many theoretical problems in statistical mechanics. We
approximate kagome lattice by Husimi lattice which is  a Bethe-type
recursive lattice. This approach allows us to plot magnetization
curves for different values of temperature and exchange parameters.
For the finite size lattice at low temperature there are bifurcation
points near plateaus, therefore Lyapunov
exponent\cite{lyapunov,lyapunov1,lyapunov2,lyapunov3} of recursion
relation are studied to verify disappearance of the bifurcation
points in thermodynamic limit.

In the strong external magnetic field   Heisenberg model can be
approximated by  Ising one. There are not reasonable conditions in
solid and fluid $^3He$ under which we could neglect the non-diagonal
Heisenberg interaction, nevertheless,  if  the strong magnetic field
is directed along the z-axis, we expect that it reduces the
transverse fluctuations. It is supposed that in this case $\sigma^x$
and $\sigma^y$-spin components are infinitely small and these spin
components can be
neglected\cite{Roger,isingpotshezenberg1,isingpotshezenberg3}.

The knowledge  of  the partition function  is very important in
statistical mechanics, since the thermodynamic functions of the
system can be expressed by means of the partition function. When the
system undergoes phase transition, some thermodynamic functions
(such as free energy) become nonanalytic at that point, therefore
phase transitions can be associated with zeroes of partition
function. In 1952 for the first time Yang and
Lee\cite{yanglee,yanglee1} offer a method for studying phase
transitions by mean of the partition function zeroes. The new
concept of  the partition function zeroes on the complex magnetic
field plane was introduced (\textit{Yang-Lee zeros}). They studied
the partition function of the Ising model as a polynomial in
activity ($e^{-\frac{2H}{k_BT}}$, where $H$-is the complex magnetic
field) and proved a circle theorem which states that for Ising model
the zeroes of the partition function lays on the unit circle on the
complex activity plan. It was shown that in thermodynamic limit the
system undergoes  phase transition only when distribution of zeros
on the complex activity plan cuts the real axes.

The Yang-Lee zeros can be studied using the dynamical systems
approach\cite{dynamical,dynamica1,dynamical2} or transfer matrix
method\cite{transfr,transfr1,transfr2,transfr3}.  According
 to    Biskup \textit{et al.}\cite{biskup}
and Monroe\cite{monroe}  the Yang-Lee zeroes correspond to the
phase coexistence lines
 on the
complex magnetic field plane. Phase coexistence lines correspond
to the points where recursive function absolute derivatives in two
fixed points are attractive and equal\cite{dynamica1}.

This paper is organized in the following way. Section 2 is devoted
to  the  investigation of the two-, and  three-site exchange
interactions Heisenberg model on kagome lattice in an external
magnetic field. In Section~3, for the strong external magnetic field
Heisenberg model is approximated  by Ising-like one on Husimi
lattice and exact recursion relation for the partition function and
magnetization are derived. In Section 4 the magnetization curves for
different values of temperature and exchange parameters have been
plotted. The  absents of bifurcation points at low temperature have
been shown by studying  Lyapunov exponent. In Section 5 Yang-Lee
zeroes are studied using dynamic system approach. Finally, Section 6
contains the concluding remarks.

\section{Two-, and Three-Site Exchange Interaction Heisenberg Model for Fluid
and Solid $^3He$ on Kagome Lattice}

 The Hamiltonian for $^3He$ consists of two parts \begin{equation}
H = H_{ex}  + H_Z,
\end{equation}
where $H_{ex}$ is spin exchange interaction Hamiltonian and $H_Z$ is
the Zeeman Hamiltonian which is responsible for magnetism. In the
most general form\cite{Roger} the multiple spin exchanges
Hamiltonian can be written as\begin{equation} H_{ex}  =  -
\sum\limits_{n,\alpha } {\mathrm{J}_{n\alpha } \left( { - 1}
\right)^p P_n, }
\end{equation}
where the summation runs over all permutations of particles, $P_n$
is the permutation operator of n particles, $\mathrm{J}_{na}$ is the
corresponding exchange energy ($\alpha$ distinguishes topologically
inequivalent cycles), and $p$ is the parity as defined in
permutation group theory, i.e., it is odd (even) if the
decomposition of the permutation into a product of pair
transpositions involves an odd (even) number of transpositions. For
third layer of   $^3He$ absorbed on the surface of graphite (see
Fig. 1~(a)) with two- and three-site exchange interactions $H_{ex}$
takes the following form
\begin{equation}
\label{Hamiltonian23}
H_{ex}  = \mathrm{J}_2 \sum\limits_{Pairs} {P_{ij} }  - \mathrm{J}_3 \sum\limits_{Triangles}
{\left( {P_{ijk}  + P_{ijk}^{ - 1} } \right)},
\end{equation}
\begin{figure}[t]
\centerline{\includegraphics[width=300pt]{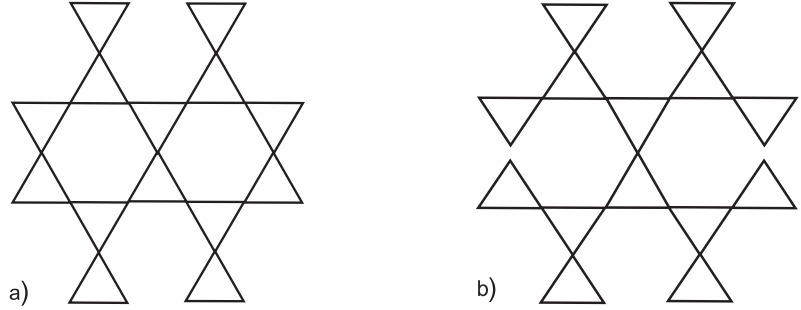}}
\vspace*{8pt} \caption{ a) kagome lattice  b) Husimi lattice}
\end{figure}where $P_{ij}$ is a pair transposition operator, $P_{ijk}$ is
an operator that
makes cyclic rearrangement in the  triangle and first sum goes over
all bonds while second sum goes over all triangles. The expression
of pair transposition operator $P_{ij}$ has been given by Dirac\cite{Dirac}
\begin{equation}\label{p2}
 P_{ij}  = \frac{1}{2}\left( {1 +  \bm{\sigma} _i
\bm{\sigma} _j } \right) ,\end{equation} where $\bm{\sigma} _i$ is
the Pauli matrix, acting on the spin at the $i$-th site. Using
Eq.(\ref{p2}) one can derive expression for $P_{ijk}$
\begin{eqnarray}
P_{ijk}=P_{ij}P_{ik} & =& \frac{1}{4}\left( {1 +  \bm{\sigma} _i
\bm{\sigma} _j } \right) \left( {1 +  \bm{\sigma} _i
\bm{\sigma} _k } \right)=\frac{1}{4}(1+\bm{\sigma} _i
\bm{\sigma} _j +\bm{\sigma} _i
\bm{\sigma} _k +(\bm{\sigma} _i
\bm{\sigma} _j)(\bm{\sigma} _i
\bm{\sigma} _k)),\nonumber\\
P^{-1}_{ijk}=P_{ik}P_{ij} & =& \frac{1}{4}\left( {1 +  \bm{\sigma} _i
\bm{\sigma} _k } \right) \left( {1 +  \bm{\sigma} _i
\bm{\sigma} _j } \right)=\frac{1}{4}(1+\bm{\sigma} _i
\bm{\sigma} _j +\bm{\sigma} _i
\bm{\sigma} _k +(\bm{\sigma} _i
\bm{\sigma} _k)(\bm{\sigma} _i
\bm{\sigma} _j)).\nonumber\\
\end{eqnarray}
Using identities
\begin{eqnarray}
(\bm{\sigma} _i
\bm{\sigma} _j)\cdot(\bm{\sigma} _i
\bm{\sigma} _k)&=&\bm{\sigma} _j
\bm{\sigma} _k
+\bm{\sigma} _i\left[\bm{\sigma} _j
\bm{\times\sigma} _k\right],\nonumber\\
(\bm{\sigma} _i
\bm{\sigma} _k)\cdot(\bm{\sigma} _i
\bm{\sigma} _j)&=&\bm{\sigma} _k
\bm{\sigma} _j
-\bm{\sigma} _i\left[\bm{\sigma} _j
\bm{\times\sigma} _k\right]
,\end{eqnarray}
one can write
\begin{equation}\label{p3}
P_{ijk}  + P_{ijk}^{ - 1}  = \frac{1}{2}\left( {1 + \bm{\sigma} _i
\bm{\sigma} _j   + \bm{\sigma} _j  \bm{\sigma} _k   + \bm{\sigma}
_k \bm{\sigma} _i } \right).
\end{equation}

Inserting equations  (\ref{p2}) and (\ref{p3}) into (\ref{Hamiltonian23})
  one can obtain the exchange Hamiltonian with two- and three-site exchange interactions\begin{equation}
H_{ex} = \frac{{\mathbf{J}_2 }}{2}\sum\limits_{\left\langle i,j\right\rangle} {\left( {1 +
\bm{\sigma} _i  \bm{\sigma} _j  } \right)}  - \frac{{\mathbf{J}_3
}}{2}\sum\limits_{\left\langle i,j,k\right\rangle} {\left( {1 + \bm{\sigma} _i \bm{\sigma}
_j + \bm{\sigma} _j  \bm{\sigma} _k
  + \bm{\sigma} _k  \bm{\sigma} _i  } \right)}
.\end{equation} The expression for the Zeeman Hamiltonian is
\begin{equation}\label{Hz}
H_Z  =  - \sum\limits_i {\frac{\gamma }{2}\hbar \bm{B}\bm{\sigma} _i}
,\end{equation}
where $\gamma$ -- is the gyromagnetic ratio for $^3He$ nucleus, $\bf B$ -- is the magnetic field. \\
Final expression for two - and three-site exchange interaction
Hamiltonian is
\begin{equation}\label{Hamiltonian}
H_{} = \frac{{\mathrm{J}_2 }}{2}\sum\limits_{\left\langle
i,j\right\rangle} {\left( {1 + \bm{\bm{\sigma}} _i  \bm{\sigma} _j
} \right)}  - \frac{{\mathrm{J}_3 }}{2}\sum\limits_{\left\langle
i,j,k\right\rangle} {\left( {1 + \bm{\sigma} _i  \bm{\sigma} _j +
\bm{\sigma} _j  \bm{\sigma} _k
  + \bm{\sigma} _k  \bm{\sigma} _i} \right)  - \sum\limits_i {\frac{\gamma }{2}\hbar \bm{B}\bm{\sigma} _i}}
.\end{equation}

\section{Ising Approximation of Heisenberg Model on Husimi Lattice. Recursion Relation for The Partition Function  }

Several approximations can be applied in the Heisenberg
Hamiltonian (\ref{Hamiltonian}). At first the Pauli matrices in
(\ref{Hamiltonian}) can be replaced by classical three-dimensional
vectors $\vec s$ of unit length (so called O(3) classical
Heisenberg model). Moreover, in the strong external magnetic field
aligned along the z-axis the contribution  from  x and y
components of classical spin variables will be insignificant and
the main contribution
 will be from the z component, which can effectively take values $s_z = \pm1$, so
  instead of  Heisenberg model
we have the Ising one. Consequently instead of   Hamiltonian
(\ref{Hamiltonian}) we get,\begin{equation}\label{HamiltonianIsing}
H = \frac{{\mathrm{J}_2 }}{2}\sum\limits_{\left\langle
i,j\right\rangle} {\left( {1 + s_i s_j} \right)}  -
\frac{{\mathrm{J}_3 }}{2}\sum\limits_{\left\langle
i,j,k\right\rangle} {\left( {1 + s _i
 s _j   + s _j  s _k   + s _k
  s _i  } \right)}  -\mathrm{h}\ \sum\limits_i {s _i}
,\end{equation} where    $\mathrm{h}\equiv\frac{\gamma }{2}\hbar
\bf {B}_{z}$.

The kagome lattice can be approximated by Husimi lattice which is a
recursive one (see Fig. 1). The recursive lattice  gives an
opportunity to obtain the exact  recursion relation for the
partition function and apply dynamic system theory. The recursive
Husimi lattice is formed in the following way: to each site of the
first (central) triangle $\gamma-1$ triangles are attached. The
construction of Husimi lattice continues recursively for each
triangle. If $\gamma=2$ then Husimi lattice is an approximation of
the kagome one. (see Fig.~1 (b)).

\begin{figure}[t]
 \label{husimi}
\centerline{\includegraphics[width=220pt]{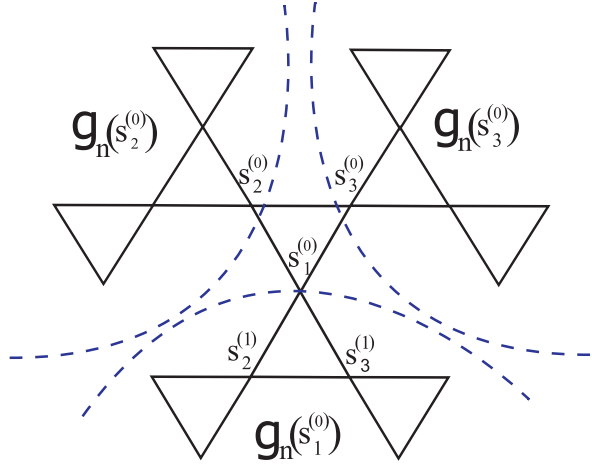}} \vspace*{8pt}
 \caption{The procedure of derivation of the recursion relation for the partition function on Husimi lattice}
\end{figure}
 Using the hierarchical structure of  Husimi lattice one can derive recursion
relation for the partition function. The partition function of the
system with Hamiltonian (\ref{Hamiltonian}) is
\begin{equation}
Z = \sum\limits_{\{ s _i \} } {e^{ -   H\left(s_1,s_2,...s_n
\right)/k_BT}} ,\end{equation} where $ k_B$ is Boltzmann constant,
$T$ is temperature  and the sum goes over all $\{s_1,s_2,...s_n\}$
configurations of the system. In the rest of the paper all
constants will be taken in Boltzmann constant's scaling
($\mathrm{J}_2/k_B \equiv J_2,\ \mathrm{J}_3/k_B \equiv J_3,\
\mathrm{h}/k_B \equiv h$).  Using Eq. (\ref{HamiltonianIsing}) one
can write
\begin{equation}\label{HamiltonianIsing1}
- \frac{H\left(s_1,s_2,...s_n \right)}{k_{b}T} =- \frac{{J_2 }}{2\,
T}\sum\limits_{\left\langle i,j\right\rangle} {\left( {1 + s_i s_j}
\right)}  + \frac{{J_3 }}{2\, T}\sum\limits_{\left\langle i,j,
k\right\rangle} {\left( {1 + s _i
 s _j   + s _j  s _k   + s _k
  s _i  } \right)}  +\frac{h}{T} \sum\limits_i {s _i}
.\end{equation} To obtain recursion relation for the partition
function one can separate  Husimi lattice into three identical parts
(branches) and at first realize summation  over all spin
configurations on each branch,  then to sum over spins of the
central triangle. The result of the summation for each branch will
only  depend on value of corresponding spin variable (see Fig. 2).
By denoting
\begin{equation}
\Delta(s _1,s_2,s _3)=- \frac{{J_2 }}{2\,}\left( {3 + s _1 s _2  + s
_2 s _3  + s _3 s _1 } \right) + \frac{{ J_3 }}{2\,}\left( {1 + s _1
s _2 + s _2 s _3  + s _3 s _1 } \right),
\end{equation}
we can rewrite the expression for the partition function in the
following form:
\begin{equation}
Z = \sum\limits_{s^{(0)} _i } {exp \left\{\frac{\Delta\left(s^{(0)}
_1,s^{(0)}_2,s^{(0)} _3\right)}{T}+ \frac{h\left(s^{(0)} _1 +
s^{(0)} _2  + s^{(0)} _3 \right)}{T}\right\}  g_n (s^{(0)} _1 )g_n
(s^{(0)} _2 )g_n (s^{(0)} _3 )} ,\end{equation} where sum goes over
central triangle,  $ g_n (s^{(0)} _i ) $ denotes the contribution of
a branch at the $s^{(0)} _i $-th site of the central triangle and
$n$ is number of generations.

At the same way one can express  $ g_n (s^{(0)} _i )$ in terms of
$ g_{n-1} (s^{(1)} _i )$ cutting it along any site of the first
generation.  Therefore
\begin{equation}
g_n (s^{(0)} _1 ) = \sum\limits_{s^{(1)} _2 s^{(1)} _3 } {exp
\left\{\frac{\Delta\left(s^{(0)} _1,s^{(1)}_2,s^{(1)} _3\right) }{T}+ \ \frac{h\left(s^{(1)} _2  + s^{(1)} _3 \right)}{T}\right\} g_{n - 1} (s^{(1)} _2
)g_{n - 1} (s^{(1)} _3 ).}
\end{equation}
After the summation we get the following expressions for
$g_n(s_i)$
\begin{eqnarray}\label{gn}\nonumber g_n ( +
) &=& e^{\frac{- 3J_2  + 2J_3  + 2h}{T}} g_{n - 1}^2 ( +
)\,\, + \,\,2e^{- \frac{J_2}{T} } g_{n - 1} ( + )g_{n - 1} ( - )\,\, +
\,\,e^{ \frac{- J_2  - 2h}{T}} g_{n - 1}^2 ( - ),\\\nonumber
g_n ( - ) &=& e^{ \frac{- J_2  + 2h}{T}} g_{n - 1}^2 ( + )\,\, + \,2e^{- \frac{J_2}{T} } g_{n - 1} ( + )g_{n - 1} ( - )\,\, + \,\,e^{ \frac{- 3J_2  + J_3  - 2h}{T}} g_{n - 1}^2 ( - ),\\
\end{eqnarray}
where we denote $g_n(s_i)$ by $g_n(\pm)$ depending on $s_i$ sign. By
introducing  $ x_n = g_n ( + )/g_n ( - ) $ variable the recursion
relation for the partition function can be obtained
\begin{eqnarray}
x_n&=&f(x_{n-1}), \label{recursive function} \nonumber\\
f(x)  &=& \frac{1 + 2cx + ac^2 x^2 }{a + 2cx  + c^2 x^2},
\end{eqnarray}
where  $a \equiv e^{\frac{2 (J_3  - J_2 )}{T}},\ c \equiv
e^{\frac{2 h}{T}} $. The thermodynamic functions of the system,
such as magnetization, can be expressed in terms of $x_n$. The
magnetization of the $s^{(0)}_1$ site is expressed
\begin{equation}\label{magnetization} m=\left\langle
s^{(0)}_1\right\rangle=\frac{\sum\limits_{\{ s _i \} } {
s^{(0)}_1e^{ - H\left(s_1,s_2,...s_n \right)/k_BT}}}{
\sum\limits_{\{ s _i \} } {e^{ -   H\left(s_1,s_2,...s_n
\right)/k_BT}}}.
\end{equation}
Let us at first  realize  the summation   over central vertex $s_1^{(0)}$and
then over all other spin variables. After inserting equations (\ref{gn}) into (\ref{magnetization}) the formula for   magnetization  will take the following form
\begin{equation}\label{magnit}
m=\frac{{e^{ \frac{h}{T}} g_n^2 ( + ) - e^{ -\frac{  h}{T}} g_n^2 ( - )}}{{e^{ \frac{h}{T}} g_n^2 ( + ) + e^{ -\frac{  h}{T}} g_n^2 ( - )}} = \frac{{c
x_n^2  - 1}}{{cx_n^2  + 1}}
.\end{equation}
   For arbitrary value
of magnetic filed $h$, with given temperature and exchange
parameters one can draw the dependence of magnetization from
external magnetic field by implementing the simple iteration from
the recursion relation for $f(x)$, beginning with some initial
value of $x_0$,  The thermodynamic limit correspond to  infinite
number of iterations $(n\rightarrow\infty)$.

 \section{The Magnetization and Lyapunov Exponent}
 In this section the magnetic properties of the model in an external magnetic field have been studied. As known,
 $J_2$ and $J_3$ are not obtainable
in the experimental measurements, because $n$-spin exchange also
makes a contribution to  $(n-1)$-spin exchanges, but there is some
effective exchange parameter, $J=J_2-2J_3$  which  can be directly
obtained from the experiments. The magnetic properties of the model
depend on value of  effective exchange parameter and vary from
ferromagnetic into antiferromagnetic one, depending on whether two-
or three-spin exchange interaction is dominant.  From experimental
measurements and theoretical calculations we know that at
low-density region three-site exchange interaction on the regular
triangular lattice is dominant. The corresponding estimated
value\cite{experemental2} of exchange parameter is
$J=J_2-2J_3=-3.07mK$.\begin{figure}[t]
 \label{magnetization1}
\centerline{\includegraphics[width=\textwidth]{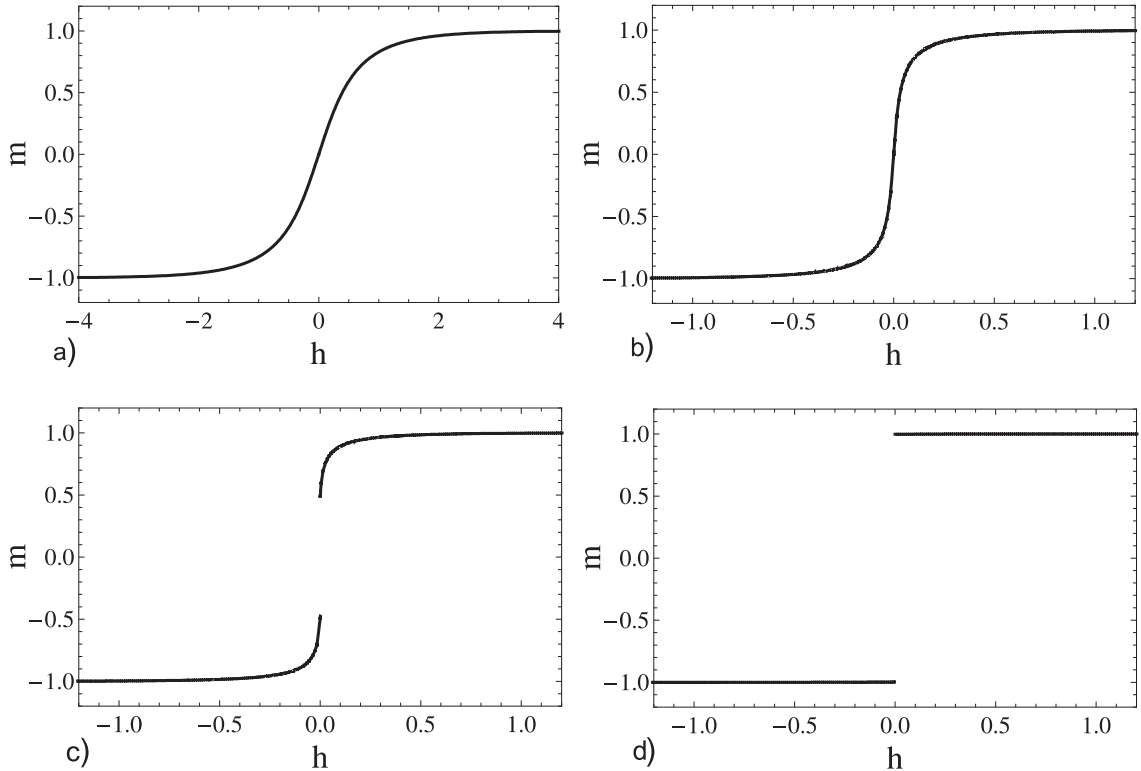}}
\vspace*{8pt} \caption{The magnetization curves for  $J_2=2mK,
J_3=2.5mK $, $n=50000$ and  for (a) $T=1.5mK$, (b) $T=0.7mK$, (c)
$T=0.6mK$, (d) $T=0.3mK$}
\end{figure}
 \begin{figure}[t]
 \label{magnetization2}
\centerline{\includegraphics[width=\textwidth]{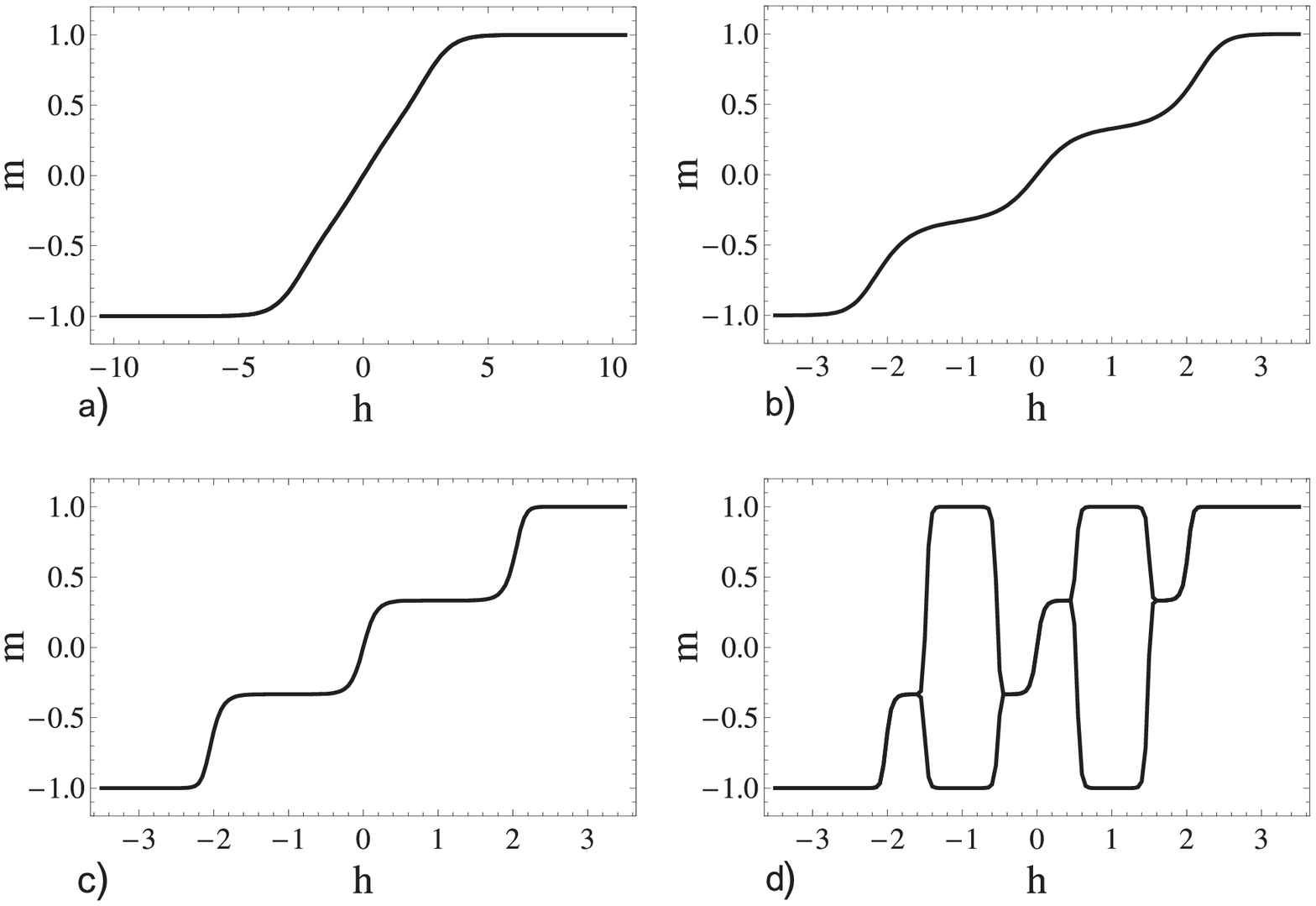}}
\vspace*{8pt} \caption{The magnetization curves for  $J_2=3mK$,
$J_3=2mK$, $N=50000$ and  for (a) $T=1mK$, (b) $T=0.3mK$, (c)
$T=0.1mK$, (d) $T=0.05mK$}
\end{figure}According to general principles ground state of the system in this case is ferromagnetic. Magnetization function of the model with exchange parameters $J_2 = 2 \ mK$
and $J_3 =2.5\ mK$ is presented in Fig. 3. For high temperatures the
magnetization curve has  monotone form of Langevin type (Fig. 3(a)).
The value of saturation field as large as higher temperature (Fig.
3(b)). For some value temperature (critical temperature) the
magnetization curve ceases to be smooth, the jump of magnetization
takes place  at an arbitrary low value of the applied magnetic field
(Fig. 3(c)). The ground state of the model  ordered ferromagnetic.
The magnetization in zero field is equal to its maximal value, which
corresponds to the ferromagnetic phase with all spins pointed in the
same direction. Under the effect of an arbitrary weak magnetic field
the twofold degeneracy of this phase is removed by orienting all
spins along the field. The corresponding diagram is presented in
Fig. 3(d).

The  model is antiferromagnetic if two-site  exchange interaction is
dominant. In Fig. 4 are presented magnetization curves that correspond
to the values of exchange parameters $J_2 = 3 \ mK$ and $J_3 =2\
mK$. As expected the magnetization curve is of Langevin type for the
high temperatures (Fig. 4(a)) and the saturation field decrease with
the temperature (Fig. 4(b)).  With  further decreasing of the
temperature plateau at 1/3 of saturation field appears on the magnetization curve (Fig.~4(c)).
This phenomena can be explained in the following way. For each
triangle two spins directed along magnetic field and the other one
oriented opposite to the field ("up-up-down"phase).

At  lower temperatures the magnetization plateau turns into
bifurcation points and period doubling (Fig. 4(d)). These
bifurcation points will disappear if  number of iterations tends to
infinity.  As it is known Lyapunov
exponent\cite{lyapunov,lyapunov1,lyapunov2,lyapunov3} is exactly
zero simultaneously with bifurcation points, indicated in some
external parameters and temperature.  The Lyapunov exponent is the
index of exponential divergence of two near points after $N$
iterations:
\begin{equation}\label{lyapunovexp} \varepsilon
e^{N\lambda (x_0 )}  \approx \left\vert f^{(N)} (x_0  +
\varepsilon ) - f^{(N)} (x_0 )\right\vert,
\end{equation}
where $f^{(N)}(x)=\overbrace {f(f( \ldots f(x)))}^{N}$.
In the limit at $\varepsilon  \to 0$ and $N \to \infty $ (\ref{lyapunovexp})
gives the exact formula for $\lambda(x_0)$
\begin{equation}\label{lyap}
\lambda (x_0 ) = \mathop {\lim }\limits_{N \to \infty } \mathop
{\lim }\limits_{\varepsilon  \to 0 }
 \frac{1}{N}\ln \left| {\frac{{f^{(N)} (x_0  + \varepsilon ) - f^{(N)} (x_0 )}}{\varepsilon }} \right| =
  \mathop {\lim }\limits_{N \to \infty } \frac{1}{N}\ln \left| {\frac{{df^{(N)} (x_0 )}}{{dx_0 }}}
  \right|.
\end{equation}
Using formula for derivative of composite function
\begin{equation}
\left. {\frac{d}{{dx}}f^{(2)} (x)} \right|_{x_0 }  = f'(f(x_0
))f'(x_0 ) = f'(x_1 )f'(x_0 ),\,\,\,x_1  \equiv f(x_0
)\label{composite},
\end{equation}
one can transform Eq. (\ref{lyap}) in the following form
\begin{equation}\label{lyapunov2}
\lambda (x_0 ) = \mathop {\lim }\limits_{N \to \infty } \frac{1}{N}\ln \left| {\frac{d}{{dx_0 }}f^{(N)} (x_0 )} \right| = \mathop {\lim }\limits_{N \to \infty } \frac{1}{N}\ln \left| {\prod\limits_{i = 0}^{N - 1} {f'(x_i )} } \right|.
\end{equation}
If  Lyapunov exponent is negative, then the final state of the
system after iterations tends to stable point of stable cycle
composed of more than one points. The positive Lyapunov exponent
result in chaotic final state of the system and zero Lyapunov
exponent correspond  to  bifurcation points.
\begin{figure}[t]
 \label{ylee1}
\centerline{\includegraphics[width=\textwidth]{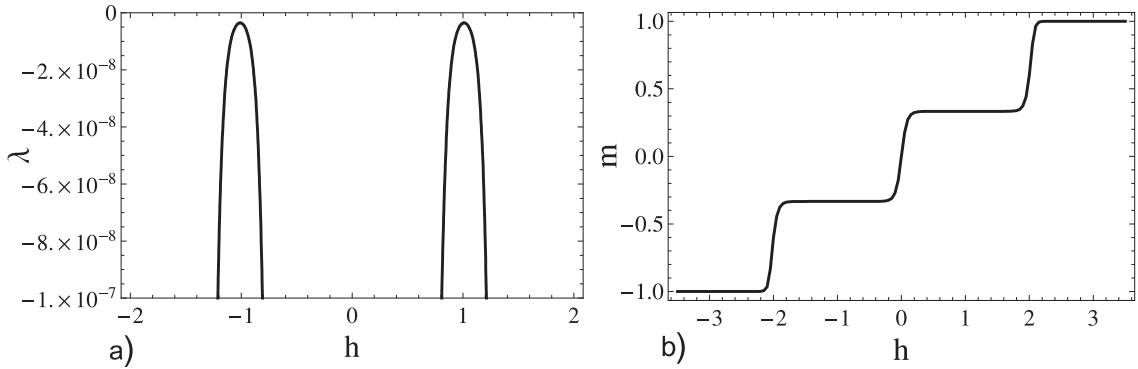}}
\vspace*{8pt} \caption{The plots of Lyapunov exponent (a) and
magnetization curve (b) for the values of exchange parameters
$J_2=3mK, J_3=2mK,$ temperature $T=0.05mK$ and iterations
n=1\,500\,000\,000. }
\end{figure}
Using Eq. (\ref{lyapunov2}) one can plot dependence of Lyapunov
exponent from external magnetic field. In Fig. 5(a) Lyapunov
exponent curve for $J_2 = 3 \ mK$, $J_3 = 2\ mK$ and $T=0.05mK$ has
been plotted. Near plateau Lyapunov exponent is very close to zero
but it does not cut the axes. In Fig. 5(b) the magnetization curve
for the same values of exchange parameters, temperature and number
of iterations $n=1\,500\,000\,000$ has been plotted. As can be seen
from figure the bifurcation points  disappeared.

\section{Yang-Lee Zeroes On Husimi lattice}

The thermodynamic properties of  the system  may be investigated by studying the dynamics
of the corresponding recursive function (\ref{recursive function}). As known
from dynamical system theory when the number of iterations tends to infinity $(n\rightarrow\infty)
$   the resulting point can have different behavior:  1) limiting point
tends to one
 point, 2) limiting point tends to    the set of points 3) limiting point has chaotic behavior. In the first case resulting point can tend to one of the stable
fixed points. The point $x^*$ called fixed point of recursion
relation $x_n=f(x_{n-1})$ if $x^*=f(x^*)$. The fixed point can be
stable (the iterations of any  point $x_0$ near $x^*$ tend to the
$x^*$), or not stable (the iterations of any  point $x_0$ near $x^*$
tend to move away from $x^*$). The stability of fixed point depends
on value of the derivative of $f(x)$ at that fixed point $x^*$. If
$\left\vert\lambda\right\vert<1$ where $ \lambda
\equiv~f^\prime(x^*)$ then the fixed point  called attracting, if
$\left\vert\lambda\right\vert>1$ then the fixed point called
repelling  and if  $\left\vert\lambda\right\vert=1$ the fixed point
called neutral (indifferent). The values of external parameters
(temperature, magnetic field, etc.) at which system has only one
attracting fixed point correspond to a stable paramagnetic state. If
the system has two attracting fixed points then the stable state
corresponds to the fixed point with maximum value of
$\left\vert\lambda\right\vert$, moreover the values of external
parameters at which $\left \vert \lambda_1 \right\vert=\left \vert
\lambda_2 \right\vert$
  correspond to two possible ferromagnetic
states with opposite magnetizations.

The fixed points of the recursion relation (\ref{recursive
function}) determined by following equation\begin{equation} x=
\frac{1 + 2cx  + ac^2 x^2 }{a + 2cx  + c^2 x^2} .\end{equation} In
general this equation has three complex solutions. If there are only
one attracting  fixed point then the system state is paramagnetic
and phase transition does not present. The values of external
parameters
 at which system has two attracting fixed
points correspond to metastable region\cite{dynamica1}. The boundary of metastable region
can be found from the condition that one of the fixed points become neutral.
There are not phase transition on the boundary of metastable region. The phase
coexistence lines can be found from the condition that absolute values of derivatives at the fixed points become equal\cite{biskup}. When phase coexistence line
cuts the real axis then the system undergoes  first order phase transition.
The locus of the phase coexistence
line inside metastable region correspond to the locus of Yang-Lee zeroes. Consequently, the boundary of  metastable region can be determined from the following system of equations:
\begin{equation}\label{neutral}
\left\{ \begin{array}{l}
 f(x) = x \\
 f'(x) = e^{i\varphi }  \\
 \end{array} \right.
.\end{equation}
Eliminating $x$ from this equations for given temperature
and exchange parameters one can find the equation for $c$ parameter.

\begin{figure}[t]
 \label{ylee}
\centerline{\includegraphics[width=\textwidth]{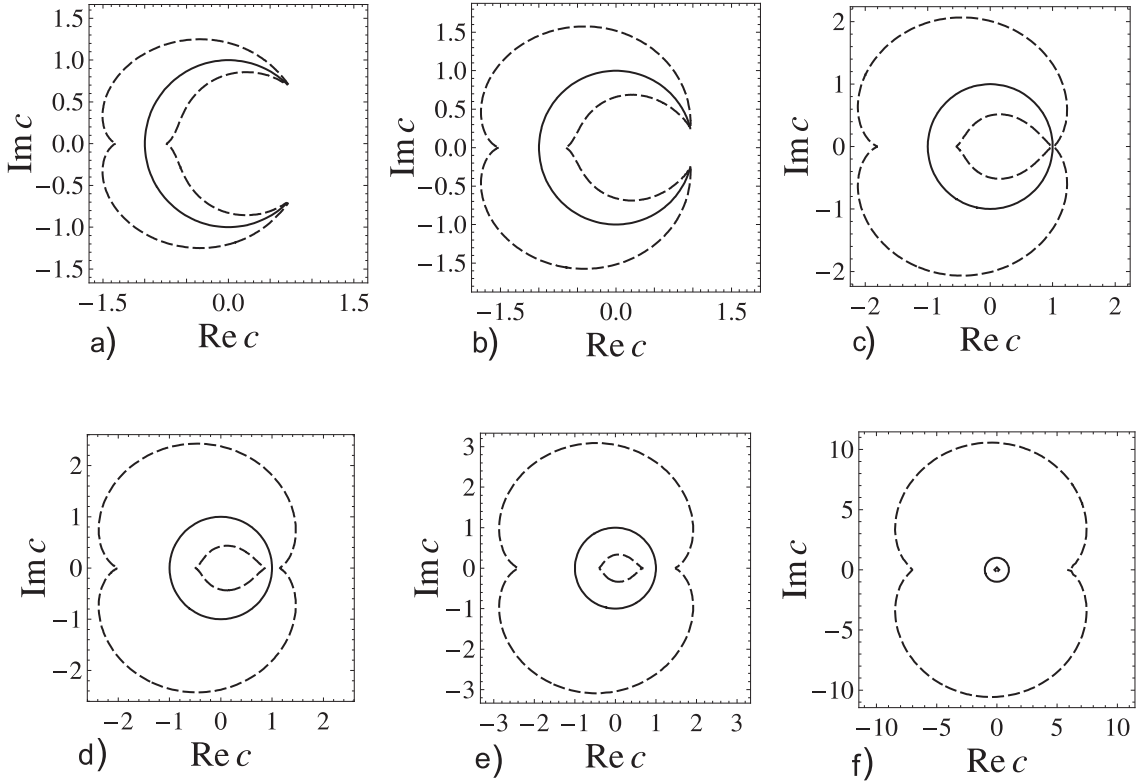}}
\vspace*{8pt}
 \caption{The border of metastable region (dashed
curve) and phase coexistence lines (solid curve) for  $J_2=2mK,
J_3=2.5mK$ and (a) $T=1.5mK$, (b) $T=0.9mK$,  (c) $T=0.6mK$, (d)
$T=0.5mK$, (e) $T=0.3mK$, (f) $T=0.1mK$.}
\end{figure}
In Fig. 6 by dashed line boundary of metastable region for $J=-3 mK$
($  J_2=2mK, J_3=2.5mK$) and different values of temperature are
plotted. By solid line phase coexistence lines inside the metastable
region are plotted . If $T>T_c$ phase coexistence line does not cut
the real axes and  is an arc of a circle with radius $R=1$ (Fig. 6
(a), (b)). For $T<T_c$ phase coexistence line cuts the real axes and
the first order phase transitions occurs, therefore system has
ferromagnetic behavior (Fig. 6 (c), (d), (e), (f)). Our calculations
show that for this values of exchange parameters
$T_c=1.2426\cdot(J_{3}-J_{2})=0,6213mK.$

\begin{figure}[t]
 \label{ylee2}
\centerline{\includegraphics[width=\textwidth]{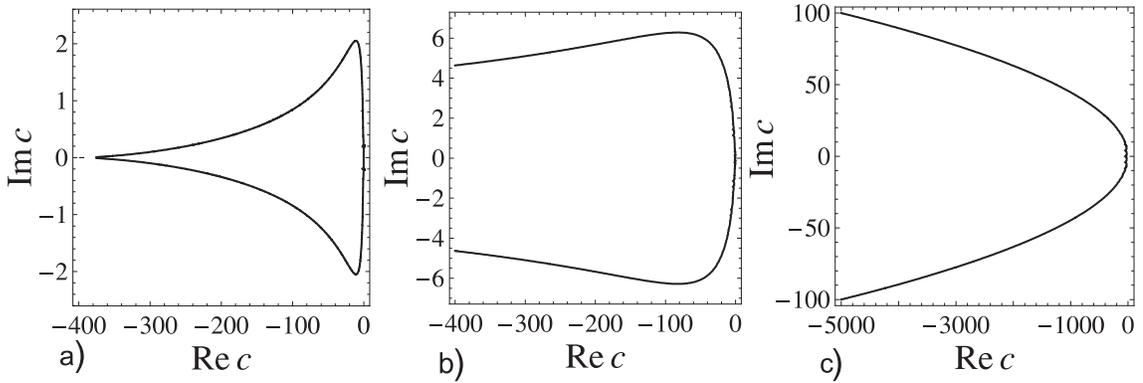}}
\vspace*{8pt}
 \caption{The border of metastable region  for  $J_2=3mK,
J_3=2mK$ and (a) $T=1mK$, (b) $T=0.5mK$, $T=0.1mK$. }
\end{figure}
If the two-site exchange interaction is dominant the system has
antiferromagnetic behavior. For this case in the metastable region
there are two attracting fixed points but they  never become equal
(no phase coexistence line in the metastable region), therefore
Yang-Lee zeroes correspond to the boundary of metastable region.  In
Figure 7 boundary of metastable region calculated using equation
(\ref{neutral}) for $J_2=3mK,J_3=2mK$ and various temperatures are
plotted. There is a relation between appearance of plateaus and
locus of the Yang-Lee zeroes.  The locus of the Yang-Lee zeroes
(Fig. 7 (a),(b)) for the values of temperature for which plateau
does not appear on magnetization curve (Fig. 4(a),(b)) have been
bounded. The locus of Yang-Lee zeroes is unbounded (Fig. 7(c)) if
the plateau appear on the magnetization curve (Fig. 4(c)).

\section{Conclusions}
In the present paper dynamic system theory has been used to study
solid and fluid $^3He$ films absorbed on the surface of graphite.
The third layer of $^3He$ films, which is kagome lattice was
approximated by $(\gamma=2)$ Husimi one. In the strong external
magnetic field the Ising model have been considered instead of
Heisenberg one with two- and three-site exchange interactions. This
approach allows us to obtain magnetization curves for different
temperatures. The magnetization plateau at 1/3 of saturation field
has been obtained for some values of exchange parameters and
temperatures. By studying Lyapunov exponent for antiferromagnetic
case the absence of the bifurcation points on magnetization plateau
at low temperature have been shown.

The Yang-Lee zeroes have been studied in terms of neutral fixed
points of the recursion relation. It was shown that Yang-Lee zeroes
on the complex $e^{2h/T}$ plane located inside the  metastable
region (existence of two  attracting fixed points)  and correspond
to phase coexistence lines (the lines where absolute derivatives  in
fixed points are equal). The locus of the Yang-Lee zeroes for
different values of exchange parameters and temperature have been
plotted. It was shown that for the ferromagnetic case Yang-Lee
zeroes are located on the arc of the unit circle. For
antiferromagnetic case if there are magnetization plateaus the locus
of Yang-Lee zeroes are infinite, otherwise they are finite.

\section*{Acknowledgment}
This work was partly supported by 1981-PS, 1518-PS,2497-PS ANSEF and
ECSP-09-08-SAS NFSAT research grants.

%+Bibliography

%-Bibliography

\end{document}